\documentclass[a4paper,twocolumn,final]{raa}

\usepackage{graphicx,times}             
\usepackage{natbib}
\usepackage{amssymb,amsmath}
\bibpunct{(}{)}{;}{a}{}{,}

\usepackage{hyperref}
\hypersetup{colorlinks = true, linkcolor = blue, anchorcolor = red, citecolor = blue, filecolor = red, pagecolor = red, urlcolor = red}

\newcommand{\mpch}{h^{-1} {\rm Mpc}}

\newcommand{\Msun}{M_{\odot}}

\newcommand{\bmc}{{\mathbf{C}}}
\newcommand{\bmcinv}{{[\mathbf{C}^{-1}]}}

\begin{document}

   \title{Weak lensing magnification reconstruction with the modified internal linear combination method
}

   \volnopage{ {\bf RAA 2021} Vol.{\bf 21} No.{\bf 10}, 247(10pp)\ \ doi: 10.1088/1674-14527/21/10/247}      
   \setcounter{page}{1}          

   \author{Shutong Hou
      \inst{1}
   \and Yu Yu
      \inst{1,2}
   \and Pengjie Zhang
      \inst{1,2,3}
   }

   \institute{Department of Astronomy, School of Physics and Astronomy, Shanghai Jiao Tong University, Shanghai 200240, China {\it yuyu22@sjtu.edu.cn}\\
        \and
             Shanghai Key Laboratory for Particle Physics and Cosmology, Shanghai Jiao Tong University,  Shanghai 200240, China\\
        \and
             Tsung-Dao Lee Institute, Shanghai, 200240,  China\\
\vs\no
   {\small Received~~2021 March 29; accepted~~2021~~June 15}}

\abstract{ 
Measuring weak lensing cosmic magnification signal is very challenging due to the overwhelming intrinsic clustering in the observed galaxy distribution.
In this paper, we modify the Internal Linear Combination (ILC) method to reconstruct the lensing signal with an extra constraint to suppress the intrinsic clustering. 
To quantify the performance,
we construct a realistic galaxy catalogue for the LSST-like photometric survey, covering $20\,000\ \deg^2$ with mean source redshift at $z_s\sim 1$.
We find that the reconstruction performance depends on the width of the photo-z bin we choose.
Due to the correlation between the lensing signal and the source galaxy distribution, the derived signal has smaller systematic bias but larger statistical uncertainty for a narrower photo-z bin.
We conclude that the lensing signal reconstruction with the Modified ILC method is unbiased with a statistical uncertainty $<5\%$ for bin width $\Delta z^P = 0.2$.
}

   \authorrunning{Shutong Hou, Yu Yu \& Pengjie Zhang }            
   \titlerunning{Weak Lensing Reconstruction by the Modified Internal Linear Combination Method}  

   \maketitle
%
%
\section{Introduction}           
\label{sect:intro}
Weak lensing is a powerful probe of large-scale structure and geometry of the universe (\citealt{2001PhR...340..291B, 2015RPPh...78h6901K}). It contains tremendous information on dark energy, dark matter and gravity at cosmological scales.

The most successful method to measure weak gravitational lensing is through cosmic shear (\citealt{2000MNRAS.318..625B,2008ARNPS..58...99H}).  With the coming of large and deep surveys, such as Euclid, Large Synoptic Survey Telescope (LSST) and the Square Kilometre Array (SKA), cosmic shear measurement has great promise to improve. However, weak lensing measurement through cosmic shear suffers from systematic errors associated with galaxy shape, such as the point spread function (\citealt{2007ApJS..172..203R}) in the measurement, galaxy intrinsic alignment contamination (\citealt{2009ApJ...694..214O, 2015PhR...558....1T}), and etc.
Only galaxies with good shape measurement are used in the cosmic shear analysis.
  
Another way of lensing measurement is through cosmic magnification. Cosmic magnification is the lensing-induced changes in galaxy number density. On one hand, lensing effect increases (decreases) the flux of sources and more (less) sources can be detected. On the other hand, cosmic magnification enlarges (reduces) the solid angle of the sky patches. The galaxy number density after lensing suffers from the two competing effects and depends on the the logarithmic slope of source galaxy luminosity function. In ideal case, the cosmological information extracted from cosmic magnification two point statistics is equivalent to that from cosmic shear. The measurement of cosmic magnification does not require a good estimation of galaxy shapes because we simply count the galaxy number and therefore is free of the systematic errors that cosmic shear suffers much.

However, cosmic magnification signal is overwhelmed by the galaxy intrinsic clustering, which is the correlation of galaxy number density from the evolution of the Universe. 
Galaxy intrinsic clustering (noise) is the main obstruction to measure weak lensing by cosmic magnification. 
In order to improve signal-to-noise ratio, very massive objects are used as lenses, such as luminous red galaxies (LRGs) and clusters (\citealt{1995ApJ...438...49B},\citealt{2016MNRAS.457.3050C}), and high redshift objects are used as sources, such as Lyman break galaxies (\citealt{2009A&A...507..683H},\citealt{2012MNRAS.426.2489M}), and quasars (\citealt{1993A&A...268....1B},\citealt{2005ApJ...633..589S}) . In this way, cosmic magnification is measured through cross correlations of two spatially well separated samples.
In addition to this, the magnification effects have been detected through the shift in number count (\citealt{1979ApJ...227...30S},\citealt{2018MNRAS.476.1071G}), magnitude (\citealt{2010MNRAS.405.1025M}), flux (\citealt{2011MNRAS.411.2113J}), and size (\citealt{2014ApJ...780L..16H}). 

The application of cross-correlation is limited by the unknown foreground galaxy bias. The direct measurement of cosmic magnification is of great importance for the precision cosmology. \cite{2006MNRAS.367..169Z} first tried to separate the magnification bias from the intrinsic clustering in flux space since magnification and galaxy clustering differently depend on galaxy flux. Magnification bias changes sign with the increase of flux while galaxy bias is nearly unchanged with flux (\citealt{2017ApJ...845..174Y}). \cite{2011MNRAS.415.3485Y} proposed a method to reconstruct the lensing convergence map from the surface number density distribution of galaxies using this idea. \cite{2018ApJ...864...10Z} proposed an analytical blind separation method to extract the cosmic magnification by counting galaxies.


The internal linear combination (ILC) method is a sophisticated method in the cosmic microwave background analysis.
A clean map is reconstructed by co-adding the data at different fluxes with a set of weights that minimizes the variance of the map.
\cite{2003ApJS..148....1B} first proposed the ILC method to obtain maps with foregrounds and noise being suppressed as far as possible. 
The main advantage of this method is that it has no assumption about the noise.
Another advantage is that it is easy to implement and computationally fast (\citealt{2012MNRAS.419.1163B}). 

Analog to the cosmic microwave background case, the cosmic magnification signal is hidden in the galaxy maps of different bands.
The dependence on the flux is known.
The intrinsic clustering is the overwhelming unknown noise we want to remove.
The future surveys will provide huge number of galaxies with good photometry and photometric redshift estimation.
LSST will observe $\sim 10$ billion galaxies with number density of $30\ \mathrm{arcmin}^{-2}$ in $u, g, r, i, z, y$ bands (\citealt{2012arXiv1211.0310L}). 
The great survey power enables us to reconstruct the cosmic magnification signal by counting the galaxies.

In this work, we modify the ILC method and validate the reconstruction of the lensing magnification signal in the context of future LSST-like survey.
The structure of the paper is organized as follows. We describe the lensing basics and the methodology of the Modified ILC method in Section~\ref{sect:method}.
In Section~\ref{sect:data}, our simulation and the mock data are introduced .
We quantify the performance of the Modified ILC method in Section~\ref{sect:performance of ILC}. A conclusion is made in Section~\ref{sect:conclusion}.

\section{Lensing Magnification and the Modified ILC method}
\label{sect:method}

\subsection{Lensing Basics}

The foreground matter distribution deflects the light-rays, leading to the lensing effects on the galaxies.
To describe the lensing field, a common quantity is 
the dimensionless lensing convergence $\kappa$, which is the projection of the matter overdensity $\delta_m$ weighted by the lensing kernel $W_L$,
\begin{equation}
    \kappa(\vec{\theta},z_{s})=\int_{0}^{z_{s}} \delta_{m}(\vec{\theta},z_{L})W_L(z_{L},z_{s})\frac{dz_{L}}{H(z_{L})/H_{0}}\ ,
    \label{eq:convergence}
\end{equation}
\begin{equation}
    W_L(z_{L},z_{s})=\left\{
\begin{aligned}
 &\frac{3}{2}\Omega_{m}(1+z_{L})\widetilde{\chi}(z_{L})(1-\frac{\chi(z_{L})}{\chi(z_{s})})\ ,
\\&  \hspace{1.5cm} z_{L}<z_{s}\\
 &0\ ,\hspace{1.1cm} z_{L}\geq z_{s}\\
\end{aligned}
\right.\ .
\label{eq:lensing_kernel}
\end{equation}
Here, $z_{s}$ and $z_{L}$ are the redshift of source and lens.
$\chi$ is the comoving distance and $\tilde{\chi}=\chi/(c/H_0)$\ . $\Omega_{m}$ is the matter density today and $H$ is the Hubble parameter.

The lensing convergence is related to the lensing potential through $\nabla^2\psi=2\kappa$, and the deflection angle $\vec\alpha=\nabla\psi$.
All the galaxy positions are changed according to this deflection field.
We can derive the change of galaxy shape and flux by looking at the Jacobian,
\begin{equation}
    A_{ij}=
    \delta_{ij}-\frac{\partial^{2}\psi}{\partial\theta_{i}\partial\theta_{j}}=
    \left( \begin{array}{cc} 
    1-\kappa-\gamma_{1} & -\gamma_{2} \\
    -\gamma_{2}   &  1-\kappa+\gamma_{1}   
    \end{array} \right).
    \label{eq:jacobian}
\end{equation}
The diagonal part $\kappa= \frac{1}{2}\nabla^{2}_{\theta}\psi$ describes the isotropic zoom of the image, and the off-diagonal part $\gamma=\gamma_1+\mathrm{i}\gamma_2$ describes the shear of the galaxy shape. Since the lensing effect does not change the surface brightness,
the flux of the galaxy is changed by $\mu=\frac{1}{\|A\|}=\frac{1}{(1-\kappa)^{2}-|\gamma|^{2}}$\ .
In weak lensing regime, $\mu\approx 1+2\kappa$\ .

Weak lensing signal could be extracted from the coherent change of the galaxy shape, so called cosmic shear.
Here we focus on the cosmic magnification, the combined effect of the change of sky area and the galaxy flux.
Totally, the galaxy number overdensity after lensing \citep{1995A&A...298..661B,2005ApJ...633..589S} is 
\begin{equation}
    \delta^{L}_{g}=\delta_{g} + g\kappa\ . 
	\label{eq:mageffect}
\end{equation}
The prefactor
\begin{equation}
    g=2(\alpha-1)\ ,\ \ \ \alpha=-\frac{d\ln{n(F)}}{d\ln{F}}-1\ ,
    \label{eq:prefactorg}
\end{equation}
is determined by $n(F)$\ , the number of galaxies per flux interval.

The power spectrum of the observed galaxy distribution $C^L_{gg}=C_{gg}+g^2C_{\kappa\kappa}+2gC_{g\kappa}$.
The lensing signal $C_{\kappa\kappa}$ is overwhelmed by the intrinsic clustering $C_{gg}$.
Usually we use galaxies within a wide photo-z bin to ensure the good measurement of the clustering.
In this case, the cross-term cannot be ignored.
Under the Limber approximation \citep{1954ApJ...119..655L}, the angular cross-power spectrum of two fields $X$ and $Y$ is an integral of the 3D matter power spectrum along with the corresponding kernel, 
\begin{equation}
   C_{XY}(\ell)=
   \int dz \frac{H(z)}{c} W^{X}(z)W^{Y}(z)\\ 
   \frac{1}{\chi^{2}}P\left(k=\frac{\ell}{\chi};z\right)\ .
\label{eq:Limber_equation}
\end{equation}
Here, $X, Y=\kappa, g$.
For galaxy field, $W^{g}(z)=n(z)$, which is the normalised redshift distribution of the galaxy sample.
For lensing field, $W^{\kappa}(z)=\int W_L(z,z_s) n(z_s) dz_s$.

\subsection{The Modified ILC method}

Extracting the lensing signal from Equation~\ref{eq:mageffect} is challenging, since the intrinsic clustering is overwhelming. The ILC method is
based on a simple premise \citep{2007astro.ph..2198D,2008PhRvD..78b3003S}: suppose there are $N_F$ observed overdensity maps with different fluxes, adding the maps with a set of weights and the aim is to minimize the variance of the map.
First, the galaxies are divided into $N_F$ flux bins. For each bin, $\delta_{g,i}^{L}=b_{i}\delta_{m}+g_{i}\kappa$.
$b_{i}$ is the deterministic bias in the $i$th flux bin.
For brevity, $\delta_{g,i}^{L}$ is denoted as $\delta_{i}^{L}$\ .
The estimated convergence map is defined as
\begin{equation}
    \hat{\kappa}=\sum_{i=1}^{N_F}w_{i}\delta_{i}^{L}\ .
\label{eq:kappa_hat}
\end{equation}

Note that the weighting can be performed over $\delta_{g}^{L}(\vec{\theta})$ in real space and over $\delta_{g}^{L}(\vec{\ell})$ in Fourier space.
The following methodology is same but they are not equivalent to each other.
In Fourier space, it minimize the noise for each angular scale $\ell$.
In this work we only validate the performance in Fourier space.

We extend the widely used ILC method in CMB analysis to extract cosmic magnification. The weights of ILC are to minimize $\langle\hat{\kappa}^{2}\rangle$, up to the constraint
\begin{equation}
\sum_{i=1}^{N_{F}} w_{i}g_{i}=1\ .
\label{eq:condition1}
\end{equation}
However, since it does not distinguish intrinsic clustering and shot noise, we find that it leads to significant bias to the reconstruction (refer to Appendix~\ref{sec:app-a} and Fig~\ref{fig:ilc_no_restrict}). Therefore we modify the ILC method by including an extra constraint 
\begin{equation}
\sum_{i=1}^{N_{F}} w_{i}b_{i}=0\ .
\label{eq:condition2}
\end{equation}
This eliminates the intrinsic clustering in the limit of deterministic bias. 
For a rough approximation, bias $b_{i}$ is regard as a constant and Equation~\ref{eq:condition2} is reduced to
\begin{equation}
\sum_{i=1}^{N_F}w_{i}=0\ .
\label{eq:condition2x}
\end{equation}
The auto power spectrum of Equation~\ref{eq:kappa_hat} can be written as 
\begin{equation}
    \left\langle \left( \sum_{i=1}^{N_F}w_{i}\delta_{i}^{L} \right)^{2} \right\rangle =\sum_{i,j=1}^{N_F}w_{i}w_{j}C_{ij}\ .
\label{eq:cross_power}
\end{equation}
Here, $C_{ij}$ is the cross-power of two maps, and for each $\ell$ they form an $N_{F}\times N_{F}$ matrix $\bmc$.

Using the Lagrangian multiplier method, 
\begin{equation}
    \sum_{i,j=1}^{N_F}w_{i}w_{j}C_{ij}-\lambda_{1}\sum_{i=1}^{N_F}w_{i}g_{i}-\lambda_{2}\sum_{i=1}^{N_F}w_{i}=0~ ,
\end{equation}
\begin{equation}
    \left( 2\sum_{i=1}^{N_F}w_{i}C_{ij}-\lambda_{1}g_{j}-\lambda_{2} \right)\delta w_{j}=0~ .
\end{equation}
We find the solution to be
\begin{equation}
    w_{i}=\frac{1}{2} \left( \lambda_{1}\sum_{j=1}^{N_F}\bmcinv_{ij}g_{j}+\lambda_{2}\sum_{j=1}^{N_F}\bmcinv_{ij}  \right)~ .
\label{eq:weights}
\end{equation}
Here, the two Lagrangian multipliers are given by
\begin{subequations}
\begin{equation}
    \lambda_{1}=\frac{2\sum_{i,j=1}^{N_F} \bmcinv_{ij}}{(\sum_{i,j=1}^{N_F}\bmcinv_{ij}g_{i}g_{j})\cdot(\sum_{i,j=1}^{N_F} \bmcinv_{ij})-(\sum_{i,j=1}^{N_F} \bmcinv_{ij}g_{i})^{2}}~ , 
\end{equation}
\begin{equation}
    \lambda_{2}=\frac{-2\sum_{i,j=1}^{N_F} \bmcinv_{ij}g_{j}}{(\sum_{i,j=1}^{N_F}\bmcinv_{ij}g_{i}g_{j})\cdot(\sum_{i,j=1}^{N_F} \bmcinv_{ij})-(\sum_{i,j=1}^{N_F} \bmcinv_{ij}g_{i})^{2}}~ . 
\end{equation}
\label{eq:multipliers}
\end{subequations}

The Modified ILC method takes use of the prior knowledge that for each flux bin, the lensing contribution to the clustering is proportional to the lensing prefactor $g$.  The weights are constructed under this prior knowledge independent of the cosmology and noise, and are obtained from the noisy data itself.

In the Modified ILC method, we require a measurement of the prefactor $g=2(\alpha-1)$  for each flux bin.
According to the definition (Eq.~(\ref{eq:prefactorg})), $\alpha$ could be measured from the change of the galaxy counts in response to some small shift in the flux.
We numerically add an extra shift on the observed flux catalogue,  $F'=F(1+ 2\epsilon)$, and compare the number counts pre to and post to the shift.  Equivalently, $\alpha$ of the flux bin $[a,b]$ is estimated by
\begin{equation}
    \alpha=\frac{\log(\int_{a/(1+2\epsilon)}^{b/(1+2\epsilon)}n(F)dF)-\log(\int_{a}^{b}n(F)dF)}{\log(1+2\epsilon)}\ .
    \label{eq:getalpha}
\end{equation}
We find that the numerical calculation converges for $|\epsilon|<0.1$.  We take the average result from positive and negative shift $\epsilon=\pm 0.025$.

The noise term post to reconstruction $\sum_i w_i\delta_i$ will be largely suppressed due to the requirement $\sum_i w_i=0$.
However, the galaxy bias has weak flux dependence and the noise term may not be eliminated completely.
The remaining part has weak correlation with the lensing signal, and thus it serve as a systematic bias for the Modified ILC method.
\section{Simulation and the Mock Catalogue Construction}
\label{sect:data}

\subsection{Simulations}

We use simulations and mocks to validate the Modified ILC method.
Light-cone simulation covering the full Stage-IV lensing survey footprint is still challenge.
Here we construct 200 sky portions of size $10^{\circ}\times10^{\circ}$ instead to mimic the statistical power of the future survey.
We use $\Omega_{\rm{cdm}}=0.223$\ , $\Omega_{b}=0.045$\ , $n_{s}=0.968$\ , $\Omega_{\Lambda}=0.732$\ , $h=0.71$\ , and $\sigma_{8}=0.83$ as the cosmological parameters in this work.

For the source galaxies, we use \texttt{FASTPM} (\citealt{2016MNRAS.463.2273F}) to generate 200 boxes with size of $411.604\ \mpch$.
This box size corresponds to the map size $10\ \deg$ at $z=1$ in the cosmology used in this work.
The particle number is chosen to be $1536^3$ and we save the Friends-of- Friends (FoF) halos with mass above $2.86\times10^{10}\Msun/h$\ .
In the flat-sky approximation, we randomly pick three boxes with random axis as the line-of-sight direction, and stack them together to cover a sufficiently wide true galaxy distribution.
We add Gaussian photo-z scatter with $\sigma_z=0.05(1+z)$ to each galaxy.
Totally we obtain $200$ independent galaxy distributions.

To obtain mock galaxy catalogues with realistic observation properties,
we use the public \texttt{cosmoDC2} (\citealt{2019ApJS..245...26K}) mock catalogue to assign the observed galaxy flux in this validation.
CosmoDC2 catalogue covers 440 deg$^{2}$ of sky area to a redshift of $z=3$.
The mass limit is $1.25\times10^{10}\Msun/h$\ , smaller than the halo catalogue we constructed.  We list all the cosmoDC2 galaxies in redshift range $0.8<z<1.2$ and divide them into $30$ mass bins spanning $2.86\times10^{10}\Msun/h$ to $1.18\times10^{15}\Msun/h$\ .
For each halo, we randomly pick one galaxy in the above list within the corresponding mass bin, and assign the photometry data to the halo.
We use three bands, $r, i, g$ to test the reconstruction.
The flux limits are chosen as 27 to ensure a flux-limited sample similar to LSST.
The number density of the final catalogue is $\sim 12\ \mathrm{arcmin}^{-2}$   for photo-z range $0.8<z^P<1.2$.



\subsection{Lensing Effects}

The lensing maps are generated according to the theoretical lensing power spectrum at $z=1$ obtained by \texttt{CAMB} code (\citealt{2000ApJ...538..473L}) and Limber approximation.
Here we ignore the non-Gaussianity of these lensing maps and this does not impact the validation of the Modified ILC method.
In real Universe, the lensing field and a given photometric source galaxy sample has weak correlation (\citealt{2001MNRAS.326..326H,2015ApJ...803...46Y}), i.e. the overlap between $W^g$ and $W^\kappa$. 
We modify the lensing fields to introduce the correct correlation with the source galaxy distribution.
\begin{equation}
    \tilde\kappa(\vec{\ell})=r_{\kappa g}\sqrt{\frac{C_{\kappa\kappa}}{C_{gg}}}\delta_{g}(\vec{\ell})+\sqrt{1-r_{\kappa g}^{2}}\cdot\kappa(\vec{\ell})~  .
\label{eq:kappa}
\end{equation}
Here, $r_{\kappa g}^{2}\equiv \frac{C_{\kappa g}^{2}}{C_{gg}C_{\kappa\kappa}}$\ .
We measure this quantity under the Limber approximation for a given galaxy sample and fiducial cosmology.
$\delta_{g}(\vec{\ell})$ is the Fourier transform of galaxy field, 
and $\kappa$\ is the convergence map with no correlation with the galaxy sample.
Obviously, the power of $\tilde{\kappa}$ in Equation~\ref{eq:kappa} is $C_{\kappa\kappa}$ and the correlation of $\tilde\kappa$ and $\delta_g$ is $C_{\kappa g}$\ .  When $r_{\kappa g}=0$, the lensing convergence is independent of the galaxy number density, and $\tilde\kappa=\kappa$.

In the above construction, both the galaxy distributions $\delta_g$ and lensing maps $\tilde{\kappa}$ satisfy the periodic boundary condition.  
We use Fourier transform to compute the lensing deflection field and magnification field.
Since the light is deflected at the lens plane, it is an inverse problem that we need to solve for the observed galaxy position at which the inverse deflection converges with the unlensed galaxy position.
We use iteration to obtain the lensed position for each galaxy.
The galaxy flux is modified according to the magnification at the observed galaxy position.
We use $4096^2$ meshes to process the galaxy catalogue according to the resolution dependence test in the Appendix~\ref{sec:app-b} (Fig.~\ref{fig:resolution}).

\section{The Performance of the Modified ILC method}
\label{sect:performance of ILC}

\begin{figure}
   \centering
   \includegraphics[width=8cm, angle=0]{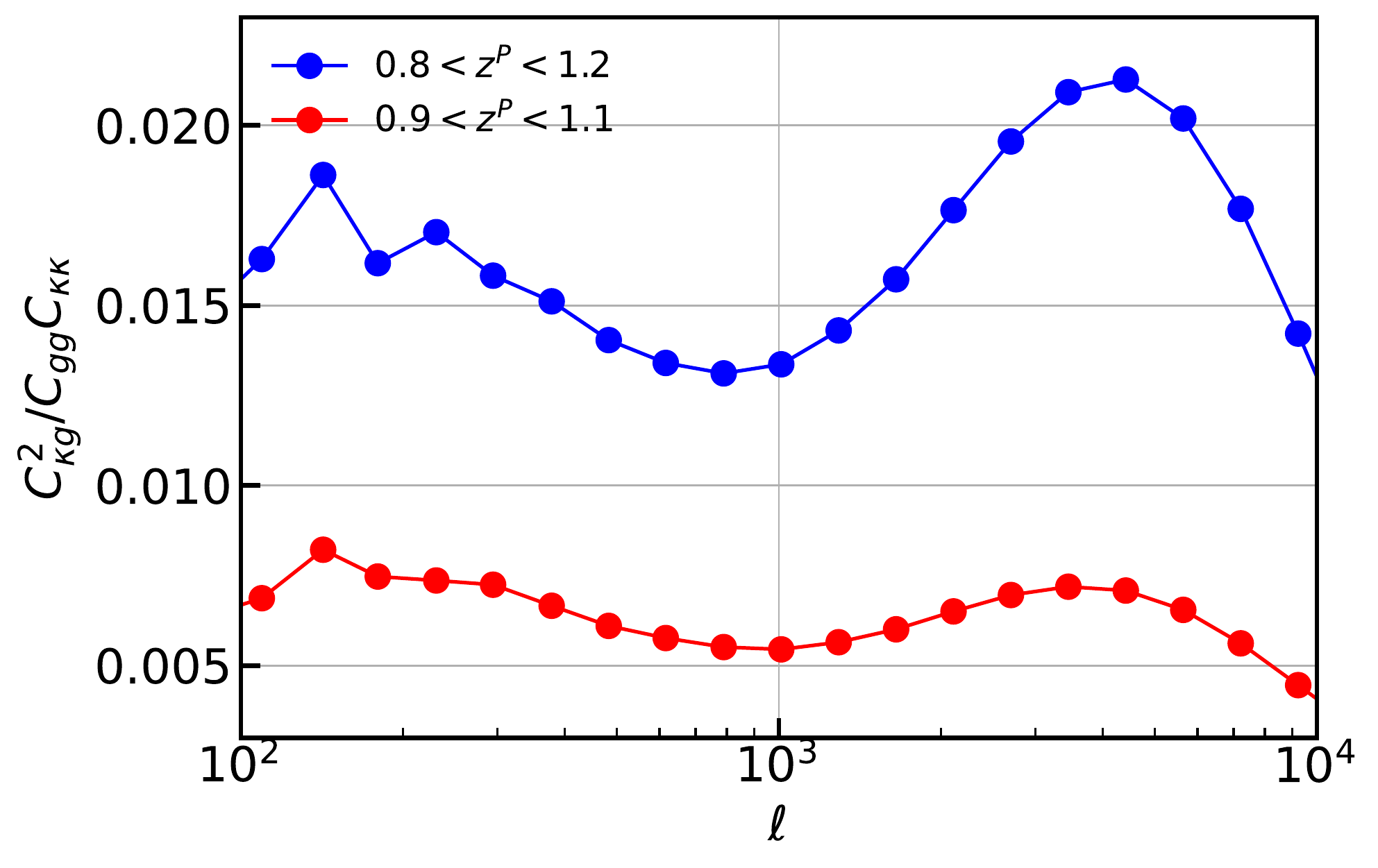}
   \caption{The cross-correlation coefficient between the distribution of the source and lensing field in the $r$-band is presented in $r_{\kappa g}^{2}\equiv \frac{C^{2}_{\kappa g}}{C_{gg}C_{\kappa\kappa}}$. The correlation is larger for the wider photo-z bin case. }
\label{fig:correlation_rband}
\end{figure}

We use the constructed lensed galaxy catalogues and the true lensing maps to test the performance of the Modified ILC method.
First, a given flux-limited photometric galaxy sample is divided into different flux bins.
The galaxy overdensity fields are measured by the CIC mass assignment and converted into Fourier space $\delta_i(\ell)$ with $i=1,\cdots,N_F$.
Then, the $g$ prefactors for each bin is numerically obtained from Eqs.~(\ref{eq:prefactorg}) and~(\ref{eq:getalpha}).
The weights $w_i$ and the Fourier space lensing field $\hat\kappa(\ell)$ is obtained by the Modified ILC method (Equations~(\ref{eq:kappa_hat}) and ~(\ref{eq:weights})).
Although the reconstructed lensing map has minimal variance by design, the shot noise is still large.
Thus, extracting the cosmological information from the reconstructed lensing field itself is challenge.
However, this noise is usually uncorrelated with the lensing measurement from other approaches such as cosmic shear.
The reconstruction is very useful to extract the cosmological information in the cross correlation with another lensing measurement.
We quantify the performance by the comparison with the true lensing map $\kappa$,
\begin{equation}
 R(\ell)=\frac{\langle\hat{\kappa}(\ell)\kappa^{*}(\ell)\rangle}{\sqrt{1-r_{\kappa g}^{2}}\langle\kappa(\ell)\kappa^{*}(\ell)\rangle}\ .
\end{equation}
Here, the cross correlation with the true lensing field in the nominator picks out the correlated part in the reconstructed lensing field.
Particularly, if the reconstruction is perfect, the nominator will be $\sqrt{1-r_{\kappa g}^{2}}C_{\kappa\kappa}$ according to Eq. \ref{eq:kappa}.
Thus, we add the prefactor $\sqrt{1-r_{\kappa g}^{2}}$ in the denominator to cancel this effect.

We test the performance for two cases.
The first one has a wide photo-z range $0.8<z^P<1.2$, while the second case has a narrow photo-z range $0.9<z^P<1.1$.
We present the cross-correlation coefficient of the convergence and galaxy number density for the $r$-band in Fig.~\ref{fig:correlation_rband}.
The wide photo-z range encloses more source galaxies in the analysis, and thus the shot noise is lower.
However, the residual intrinsic clustering is large.
$r_{\kappa g}^{2}$ is about $0.015$ at $\ell<2000$.
In the narrower photo-z range case, the number density is lower.
Meanwhile, a small cross-correlation is observed, with $r_{\kappa g}^{2}$ is about $0.009$ at $\ell<2000$.
This cross-correlation will downgrade the Modified ILC reconstruction,
and thus we expect a larger bias for the wider photo-z range.

\subsection{A Wide Photo-z Range}

\begin{figure}
   \centering
   \includegraphics[width=8cm, angle=0]{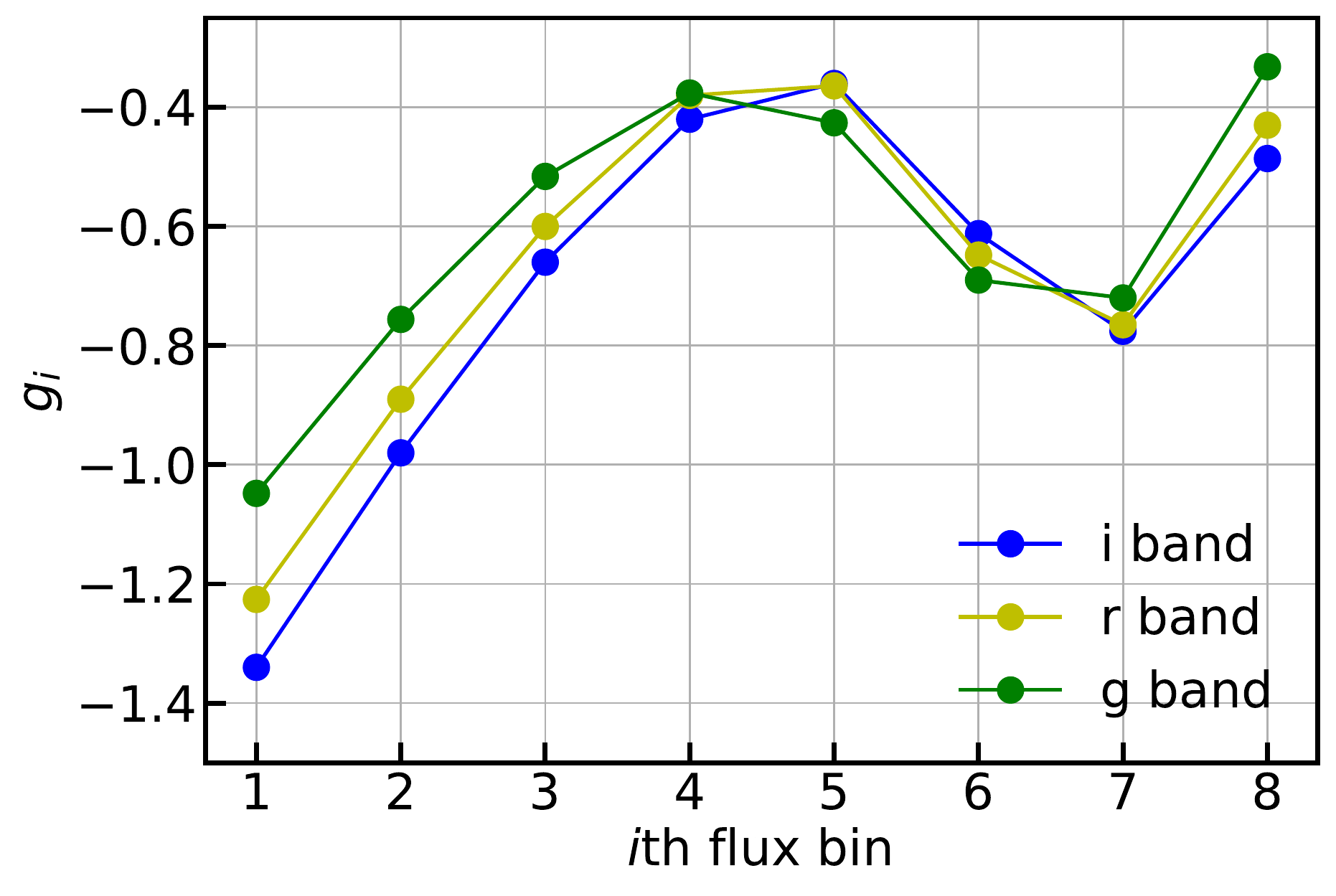}
   \caption{The $g$ prefactors on each flux bin and each band for the wide photo-z sample.}
\label{fig:gi}
\end{figure}

\begin{figure}
   \centering
   \includegraphics[width=8cm, angle=0]{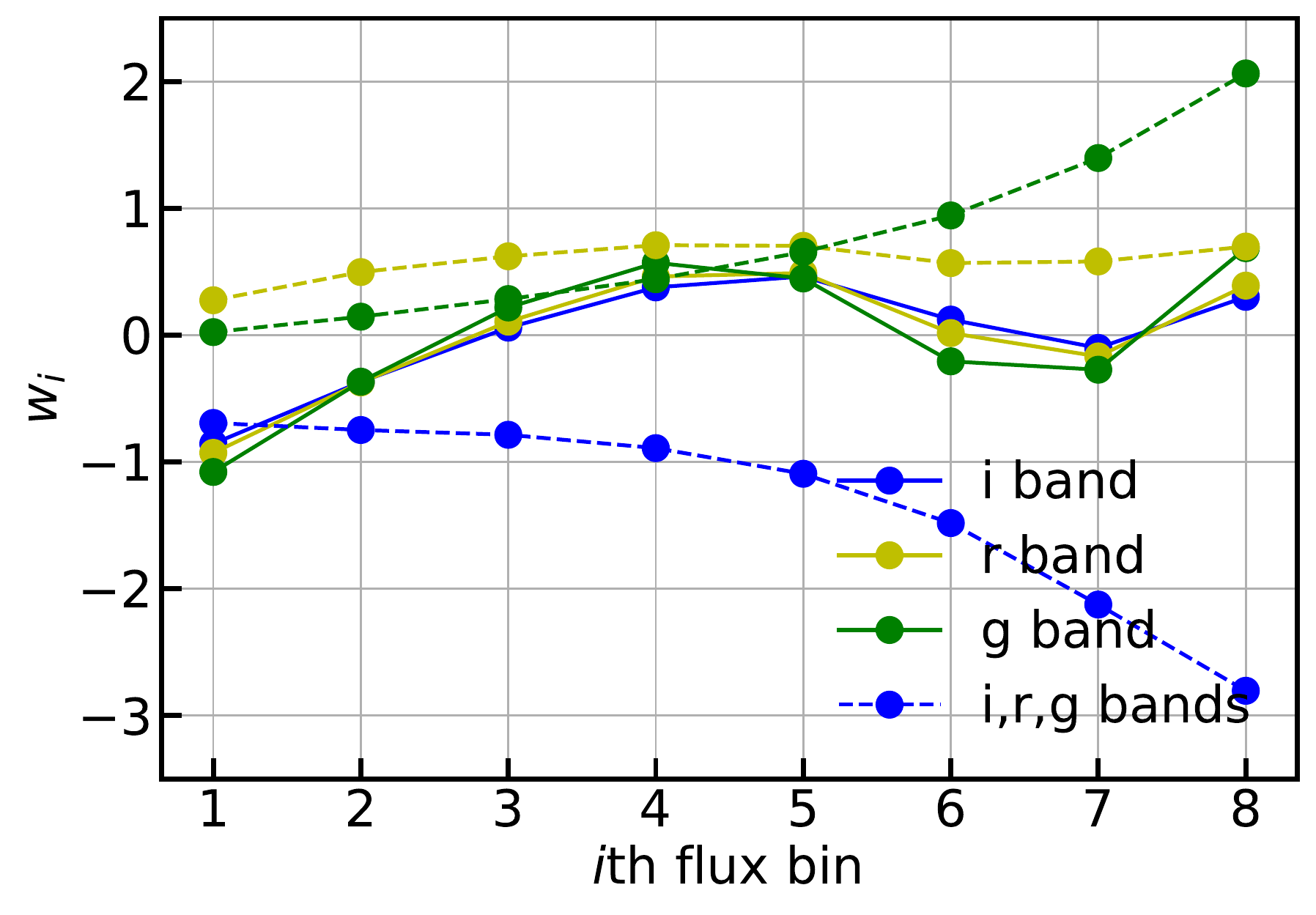}
   \caption{The weights obtained by Eq.~\ref{eq:weights} for individual bands are presented in {\it solid lines}.
   The {\it dashed lines} are the weights for the combined data with $i, r, g$ bands.}
\label{fig:weight}
\end{figure}


\begin{figure*}
   \centering
   \includegraphics[width=\textwidth, angle=0]{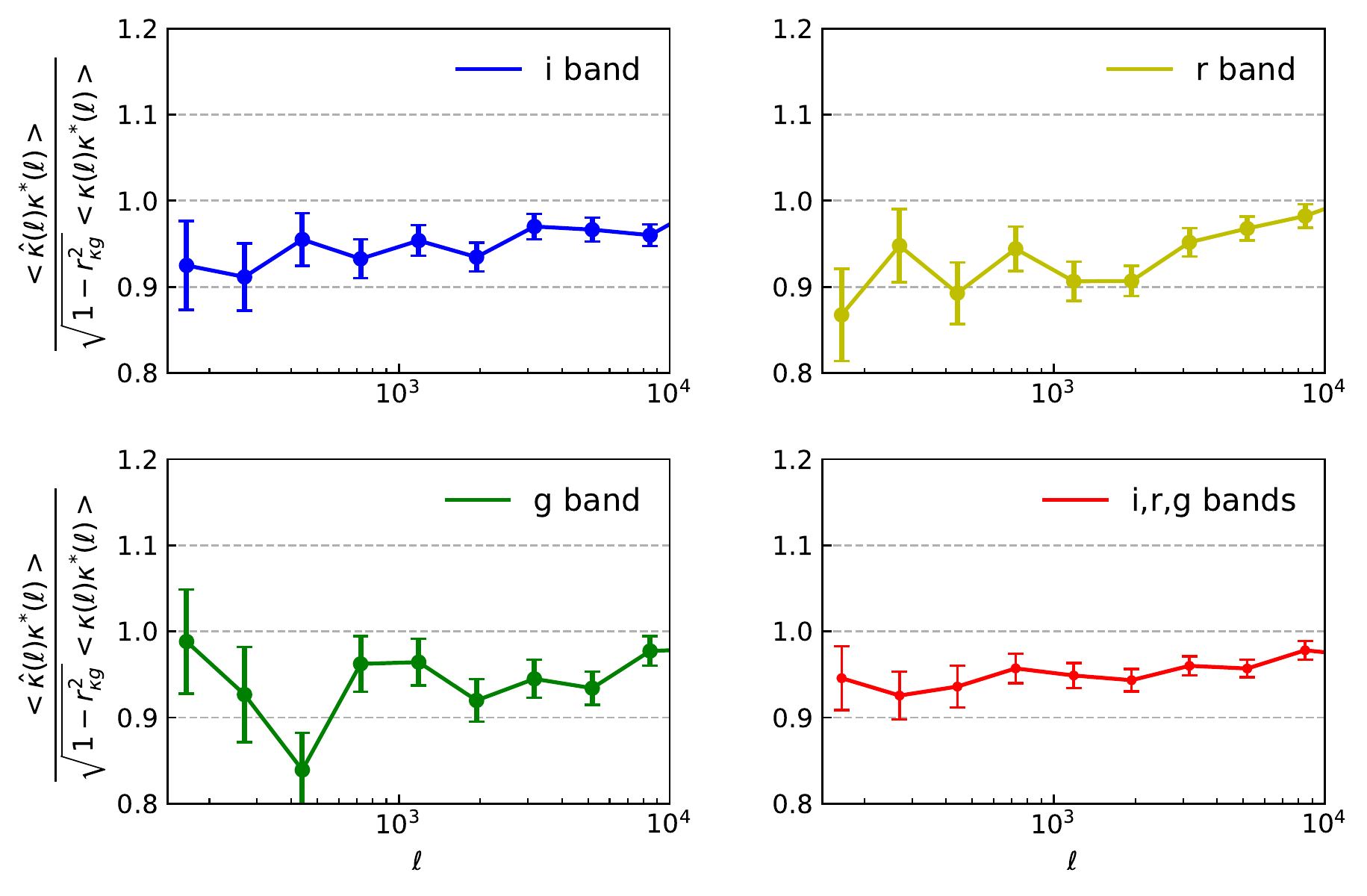}
   \caption{The Modified ILC performance is presented for three individual bands and the combined data for a wide photo-z range $0.8<z^{P}<1.2$.  The error bars are the statistical uncertainty scaled to sky coverage of $20\,000\ \mathrm{deg}^2$.  We observe $~5\%$ systematic bias in all cases due to the residual intrinsic clustering.}
\label{fig:ilc_irg1}
\end{figure*}

First we chose a wide photo-z bin range as $0.8<z^P<1.2$.
The detailed galaxy number density depends on the band, and for the $r$-band the number density reaches $12\ \rm{arcmin}^{-2}$.
For each band, we divide the galaxies into eight flux bins with the same number in each bin.
We measure the $g$ prefactors for all the bins according to Equations ~\ref{eq:prefactorg} and~\ref{eq:getalpha}.
The result is presented in Fig. ~\ref{fig:gi}.
The behavior for different bands has the same tendency with slight differences.
The galaxy overdensity is measured for all the flux bins and for all the bands.
We use $4096^2$ meshes to perform the analysis.
This resolution is sufficient to obtain reliable statistics for the range of interest.

We first test the performance for each band individually.
The resulting weights from Eq.~\ref{eq:weights} is shown in solid lines in Fig.~\ref{fig:weight}.
The weights of different bands have a similar dependence on the flux.
For each band, the lensing signal is reconstructed by adding the lensed galaxy overdensities in these flux bins with the corresponding weights.
The reconstruction performance is presented in the first three panels of Fig.~\ref{fig:ilc_irg1}.

We observe a $\sim 5\%$ underestimation of the lensing signal over the whole $\ell$ range we investigate.
The detailed number depends on the scale and band.
The error bars are calculated from the 200 sky portions and scaled to a survey with area of $20\,000\ \mathrm{deg}^2$.
The uncertainty is larger at larger scales due to cosmic variance.
The error bar reduces from $\sim 5\%$ to $1\%$ on the scales $\ell=160$ to $\ell=10\,000$.
This bias is significant compared to the statistical power.
Since the number density, the true galaxy distribution and the weights for different bands are similar, it is expected that both the systematics and the statistical uncertainty are similar for all bands.

Given that we have flux measurement on multiple bands,
a natural and optimal way is to perform the Modified ILC reconstruction on the combined data from all the bands.
Since the $g$ dependence on the flux is slightly different for each band (Fig.~\ref{fig:gi}),
we gain information from the combined sample,
and thus the reconstruction is expected to be better. 
We keep the binning scheme in each band and totally we have 24 flux bins.
The resulting weights are presented in dashed line in Fig.~\ref{fig:weight}. 
The reconstruction performance is presented in the bottom right panel of Fig.~\ref{fig:ilc_irg1}.

The performance is better when combining all bands. It has less fluctuations at different scales, and
the statistical uncertainty is smaller due to the larger number of galaxies.
However, combining the data from all bands does not alleviate the underestimation.
This also implies that the source of the underestimation is the residual intrinsic clustering.

This residual noise has weak correlation with the lensing signal.
We can reduce this correlation by simply using a narrow photo-z bin.

\subsection{A Narrow Photo-z Range}

\begin{figure*}
   \centering
   \includegraphics[width=\textwidth, angle=0]{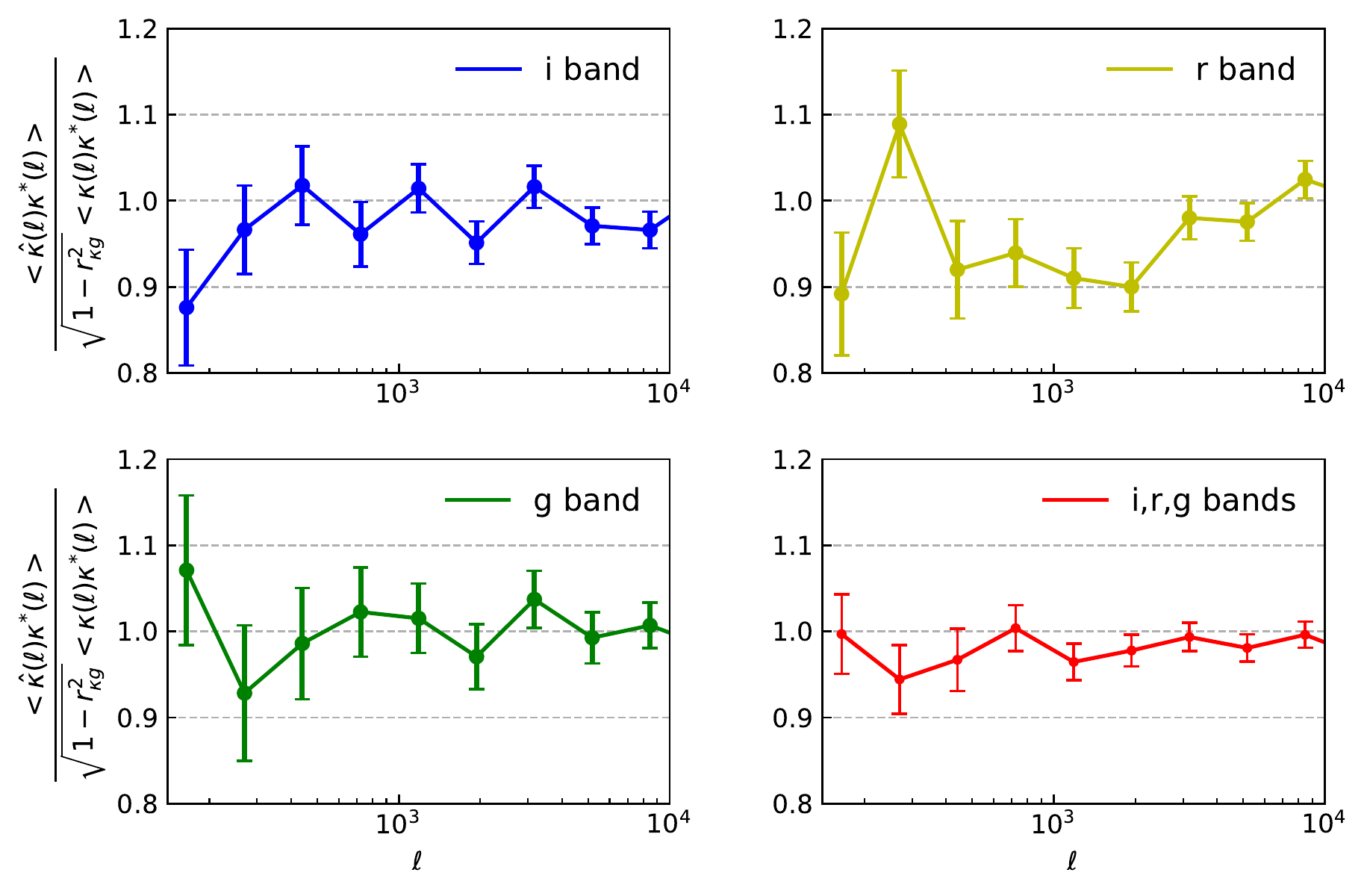}
   \caption{The Modified ILC performance for a narrow photo-z bin is presented.  The four panels show the result for three individual bands and the combined data.
   A narrow photo-z bin reduces the residual intrinsic clustering, and thus the systematic underestimation is suppressed, at the cost of slightly large statistical error.}
\label{fig:ilc_small_irg1}
\end{figure*}

The disadvantage of a narrow photo-z bin is the low number of galaxies, which increases the statistical uncertainty in the reconstruction.
The number density in this case is reduced to $\sim 6\ \rm{arcmin}^{-2}$.
As shown in Fig.~\ref{fig:correlation_rband}, $r_{\kappa g}^2$ is reduced by a factor of $2-3$.

Again, we test the performance for each band individually and for combined sample.
The result is presented in Fig.~\ref{fig:ilc_small_irg1}.
Compared with the wide photo-z bin result, we observe large fluctuations in the reconstruction performance at different scales and in different bands.
As expected, the statistical uncertainty is large in this case.
For single band, at the large scale the statistical uncertainty is about $8\%$ and for combined sample it is reduced to $5\%$.
However, there is no obvious underestimation of the lensing signal.
The deviation from 1 is within $1\sigma$ error for the combined sample. 

We present the result for the case that the galaxies are divided into four flux bins for each band in Appendix \ref{sec:app-c}. The performance is similar.

\subsection{Comparison with a Straightforward Weighting Scheme}

\begin{figure*}
   \centering
   \includegraphics[width=8.0cm, angle=0]{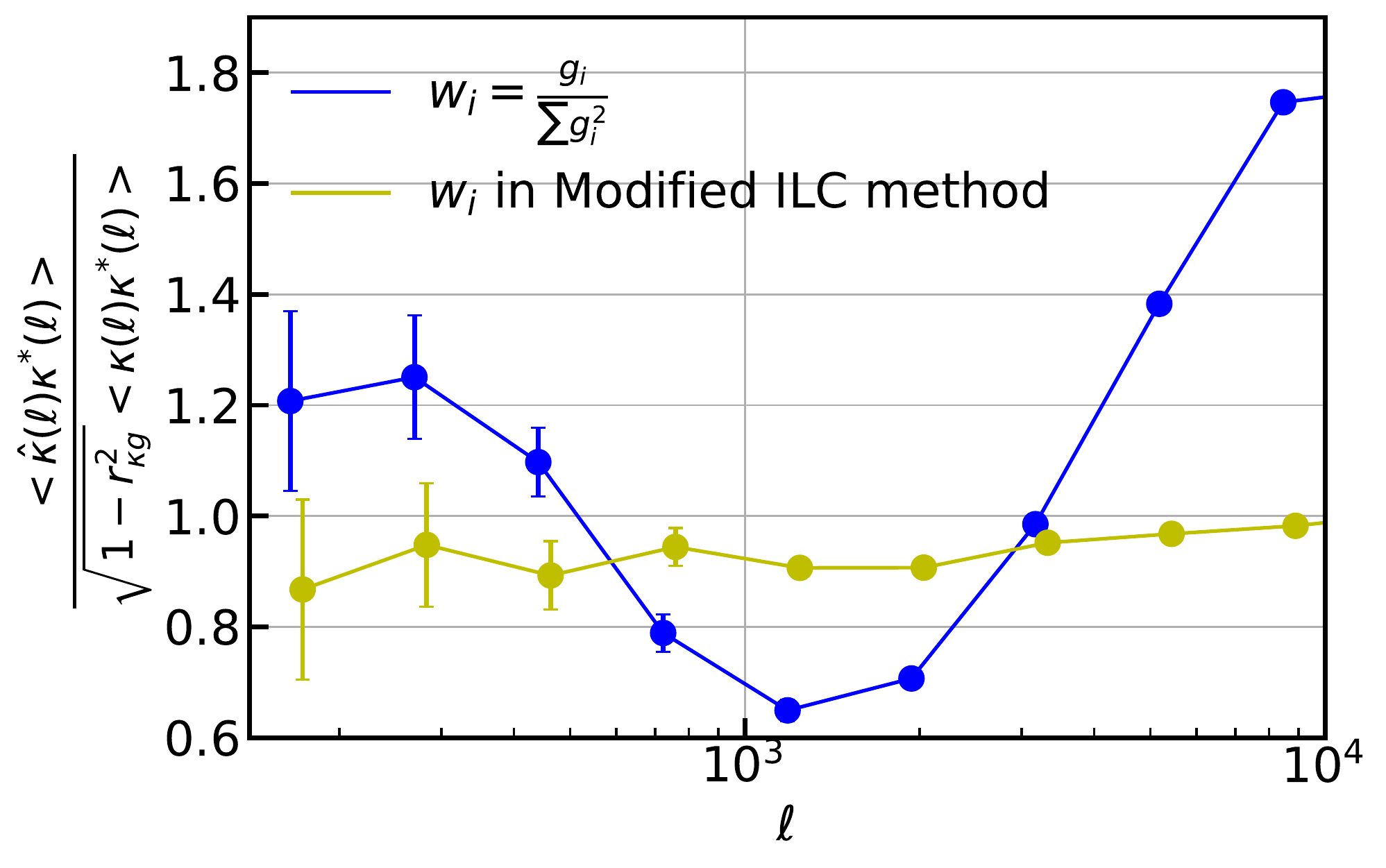}
   \includegraphics[width=8.0cm, angle=0]{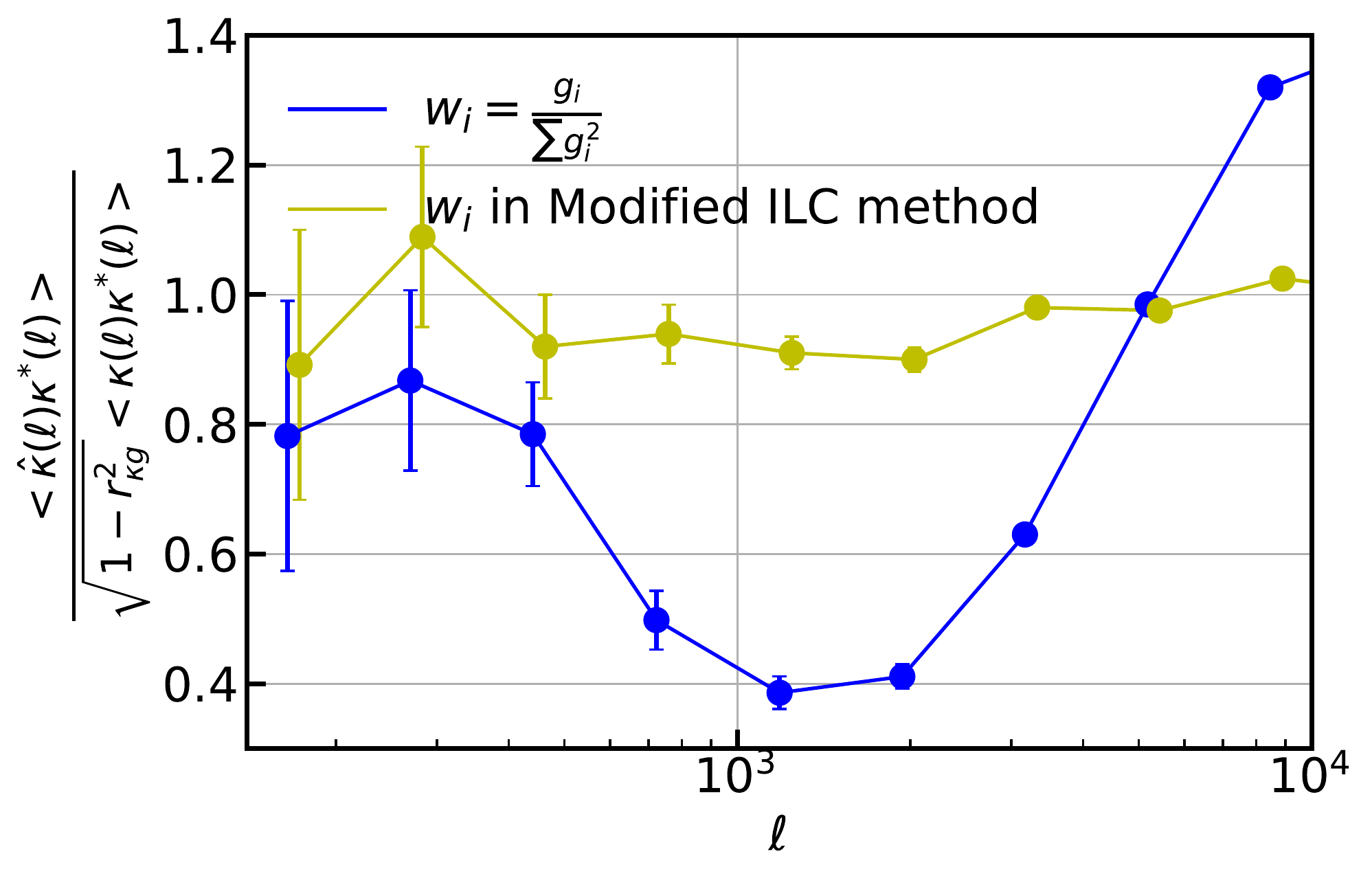}
   \caption{We compare two weighting schemes used to reconstruct the lensing convergence.  The left and right panels show the $r$-band result for $0.8<z^{P}<1.2$ and $0.9<z^{P}<1.1$. The {\it yellow lines} correspond to the Modified ILC method, and the {\it blue lines} correspond to the weighting $w_i=g_i/\sum g_i^2$.  The Modified ILC method has a much better performance.}
\label{fig:other_weight}
\end{figure*}

From $\delta_{g,i}^{L}=b_{i}\delta_{m}+g_{i}\kappa$, one can construct a straightforward weighting scheme, $w_i=g_i/\sum g_i^2$.
This weighting also has the property $\sum_i w_ig_i=1$.
Similar to the traditional ILC method, no constraint to suppress the intrinsic clustering is made.
In the case that the galaxy distribution has non-negligible correlation with the lensing signal, the residual noise $\sum_i g_i\delta_i/\sum_i g_i^2$  leads to systematics in the reconstruction.

Here we test the performance of the $w_i=g_i/\sum g_i^2$ weighting for the $r$-band in Fig. \ref{fig:other_weight}.
We observe severe systematic bias in the result for both wide photo-z bin (left panel) and narrow photo-z bin (right panel) cases.
The bias is larger for wider photo-z bin.
The residual intrinsic clustering has a large impact on this straightforward weighting scheme, even a narrow photo-z bin is adopted.
Also plotted is the result from the Modified ILC method on the same data.
The Modified ILC method has well suppressed systematics thanks to the requirement $\sum w_i=0$.

\section{Conclusion and discussion}
\label{sect:conclusion}

In this work, weak lensing magnification signal is reconstructed by the Modified ILC method on a realistic $i, r, g$ photometric mock data constructed from FASTPM simulations and cosmoDC2 mock catalogue.  The lensing fields are generated from theory and Modified to have realistic correlation with the galaxy distribution.  Then, the cosmic magnification effects are added to the galaxy sample.  We quantify the performances for two photo-z bin selections, and for cases using data from individual bands and from the combined sample.

The sample for the wide photo-z bin ($0.8<z^P<1.2$) has a number density $\sim 12\  \mathrm{arcmin}^{-2}$ for each band.
We obtain $5\%$ reconstruction uncertainty at the large scale for a survey with sky coverage $20\,000\ \mathrm{deg}^2$.
Towards small scale the statistical error decreases to $1\%$.
However, the reconstruction underestimates the lensing power at level of $\sim 5\%$.
Using the combined sample reduces the error bars but the underestimation is unchanged.
This systematic bias is the result of the correlation between the galaxy distribution and lensing signal.

Using a narrow photo-z bin ($0.9<z^P<1.1$) can reduce the correlation between the signal we want to extract and the noise we want to remove.
$r_{\kappa g}^2$ is reduced by a factor of $2-3$ depending on the scale.
We observe no obvious systematic underestimation for the cases using individual bands.
The statistical uncertainty at the large scale is increased to $\sim 8\%$ due to the low source number density ($\sim 6\ \mathrm{arcmin}^{-2}$).
Due to the same reason, the performances for individual bands show slightly large fluctuations on different scales.
The performance for the combined data is stable and has small statistical uncertainty.

In the Modified ILC method we made a rough assumption that galaxy bias is constant in all flux bins.
Future work will include the construction of new weight scheme considering the galaxy bias dependence on the flux. 
In the construction of the source galaxy catalogue, we assigned one galaxy to each halo.  The true Universe has the complexity that one massive halo can hold several galaxies and very small halo may contain no galaxy.  In other words, we approximated the galaxy bias as halo bias.  This simplification will influence the detailed performances at the small scales.  However, thanks to the extra constraint, the impact is not expected to be large.  We will use more realistic galaxy sample mock in future validation.

In this work, we use the flux limit 27 for all bands, similar to the future LSST survey. 
Including the faint galaxies decreases the shot noise and samples $g$ over a wider flux range, leading to a better performance.
We presumably will use all the available galaxies for a given flux-limited sample. 
Thus, we do not test performance dependence on the flux cuts.
In this work, galaxies are divided into flux bins with the same number of galaxies in each bin. 
This might not be an optimal choice since the brightest flux bin is very wide and the faintest bin is very narrow.
The dependence of $g$ on the flux is not well sampled.
We will explore to search for an optimal binning scheme in flux space.

In real observation, the measurement suffers the effects of survey geometry and mask.
The reconstructed lensing field is a linear combination of the galaxy density field, and thus suffers from the same effect.
One should take the mask effect in the theoretical prediction in the analysis.
On the other side, the Modified ILC method can also be developed in real space.
We leave this to the future work.
Besides the mask effect, the gray dust may change the flux of the source galaxies, and add complexities in the Modified ILC method.

\begin{acknowledgements}
This work was supported by the National Key Basic Research and Development Program of China (No. 2018YFA0404504), the National Natural Science Foundation of China (grants No. 11773048, 11621303 and 11890691), the science research grants from the China Manned Space Project (No. CMS-CSST-2021-B01),  and the "111" Project of the Ministry of Education under grant No. B20019.
\end{acknowledgements}

\appendix                  

\section{Necessity of Extra Constraint in the Reconstruction}
\label{sec:app-a}

\begin{figure}
   \centering
   \includegraphics[width=8cm, angle=0]{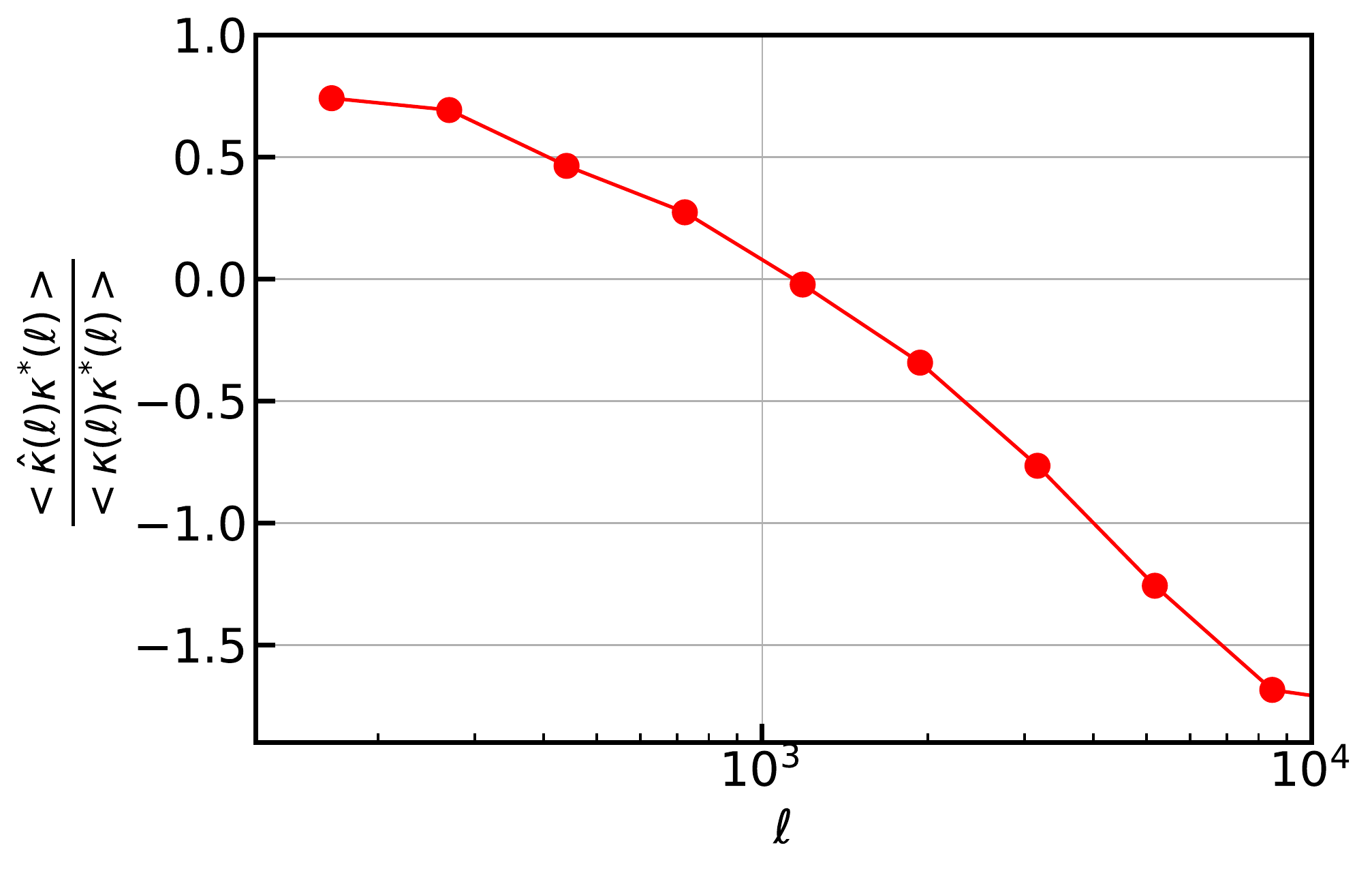}
   \caption{Reconstruction with the traditional ILC method (no constraint $\sum_{i=1}^{N_{F}}w_{i}=0$) on the $r$-band. The photo-z range is $0.8<z^{P}<1.2$. }
   \label{fig:ilc_no_restrict}
\end{figure}

Our modified ILC method includes an extra constraint $\sum_{i=1}^{N_{F}}w_{i}=0$ to suppress to intrinsic clustering. When the extra constraint is not included, the performance of reconstruction is bad as shown in Figure ~\ref{fig:ilc_no_restrict}. $\frac{\langle\hat{\kappa}(\ell)\kappa^{*}(\ell)\rangle}{\langle\kappa(\ell)\kappa^{*}(\ell)\rangle}$ decreases from $\sim+0.7$ to $\sim-1.5$ towards increasing $\ell$. The influence of intrinsic clustering on magnification reconstruction with the tradition ILC method is catastrophic and cannot be ignored. So we add the extra constraint $\sum_{i=1}^{N_{F}}w_{i}=0$.

\section{Resolution Dependence}
\label{sec:app-b}

\begin{figure}
   \centering
   \includegraphics[width=8.0cm]{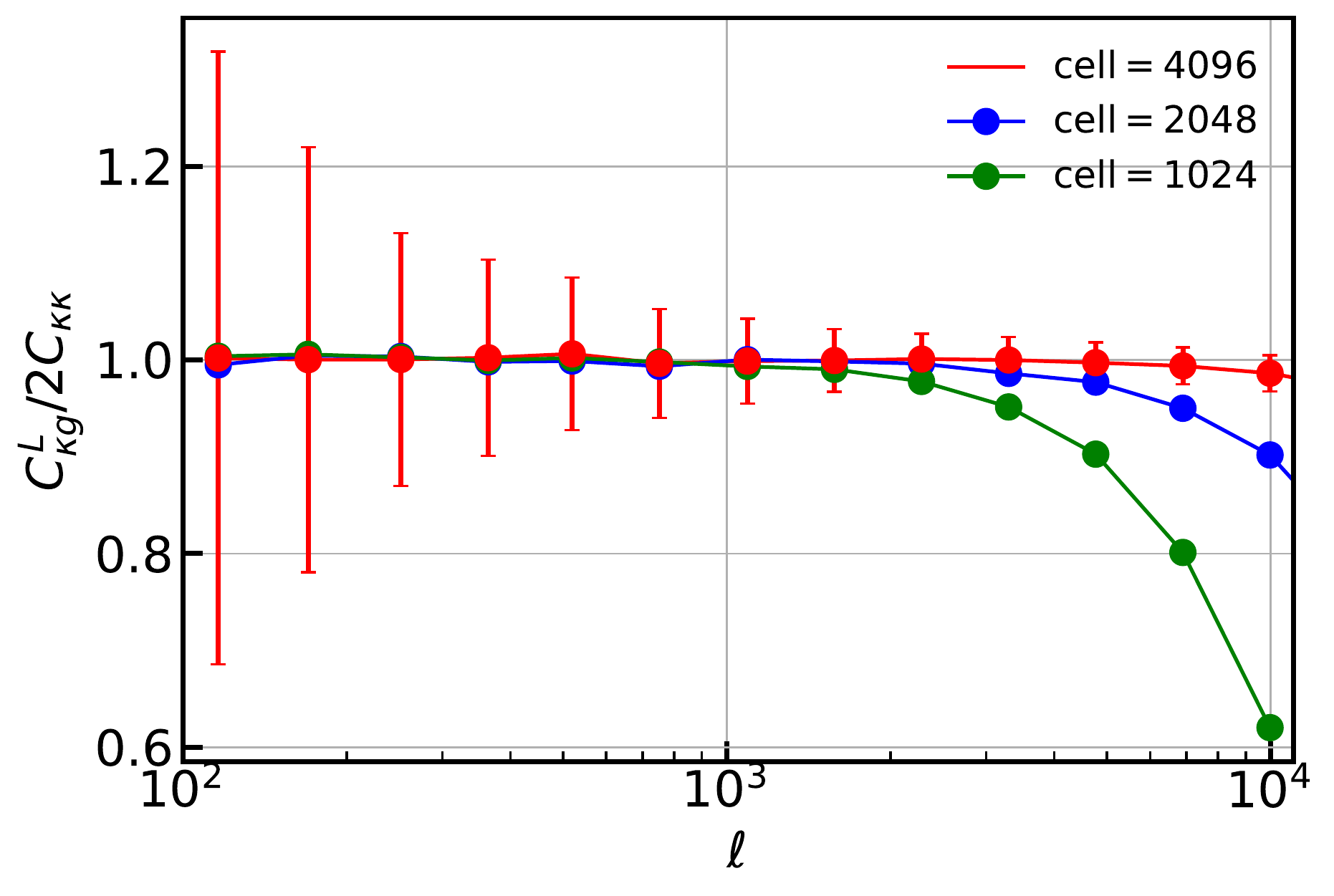}
   \includegraphics[width=8.0cm]{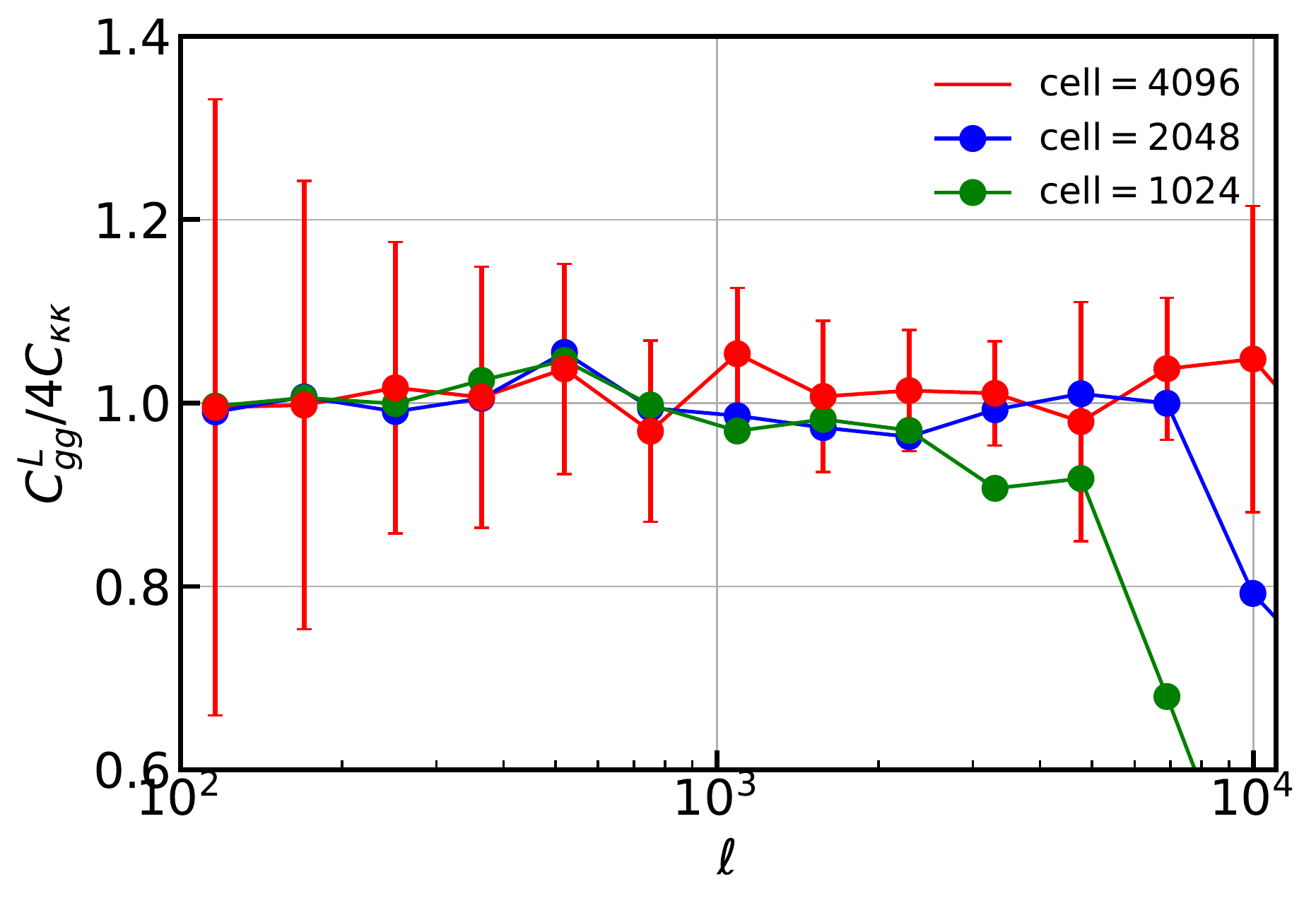}
  \caption{The left and right panels are $C^{L}_{\kappa g}/2C_{\kappa\kappa}$ and $C_{gg}^{L}/4C_{\kappa\kappa}$ used to diagnose the resolution dependence in the process of adding lensing effects.  The {\it green}, {\it blue} and {\it red lines} show the result of $1024^2$, $2048^2$ and $4096^2$ meshes, respectively.
  The result using $4096^2$ meshes shows no bias for all the scales we investigate.}
  \label{fig:resolution}
\end{figure}

We use an iterative method to solve for the lensed galaxy position where the inverse deflected position overlaps with the original galaxy position.
Interpolation is adopted to calculate the deflection angle at given positions.
In weak lensing regime, the deflection angle is small.
Thus this process is suspicious to resolution effects.
We use the following convergence test to validate the process of adding lensing effects.

We only test whether we can recover the input lensing signal from the lensed galaxy positions.  We construct an unrealistic galaxy sample, with high number density $\sim 40\ \mathrm{arcmin}^{-2}$ and no intrinsic clustering.
Lensing fields are generated on $1024^2, 2048^2$ and $4096^2$ meshes for  $10^\circ\times 10^\circ$ maps.
We solve for the lensed galaxy positions from these lensing deflection maps with different resolutions.
Theoretically, the power spectrum of the lensed sample directly related to the input lensing signal, $C_{gg}^L=4C_{\kappa\kappa}$ and $C_{\kappa g}^L=2C_{\kappa\kappa}$.

In Figure ~\ref{fig:resolution}, the left panel shows the ratio $C_{\kappa g}^L/2C_{\kappa\kappa}$ and the right panel shows $C_{gg}^L/4C_{\kappa\kappa}$.
The error bars are the r.m.s. of 200 realizations.
We find that for the two cases with low resolution, both the auto power spectrum and the cross power spectrum are underestimated at small scales.
There is no bias observed in the result using $4096^2$ meshes.
From this convergence test, we determine to use $4096^2$ meshes in the analysis.

\section{Dependence on the number of flux bins}
\label{sec:app-c}

We test the dependence on the number of flux bins used in the Modified ILC method. 
Generally, the density measurement for each flux bin becomes noisy when a large number of bins is adopted.
However, the reconstruction takes use of the measurement over all flux bins.
We do not expect a large difference as long as the choice is reasonable.
The number of bins is chosen to be sufficiently large to sample the flux dependence of the clustering, and small enough to obtain a reliable clustering measurement for each bin.
Finding the optimal binning strategy is a non-trivial task.

\begin{figure}
   \centering
   \includegraphics[width=8cm, angle=0]{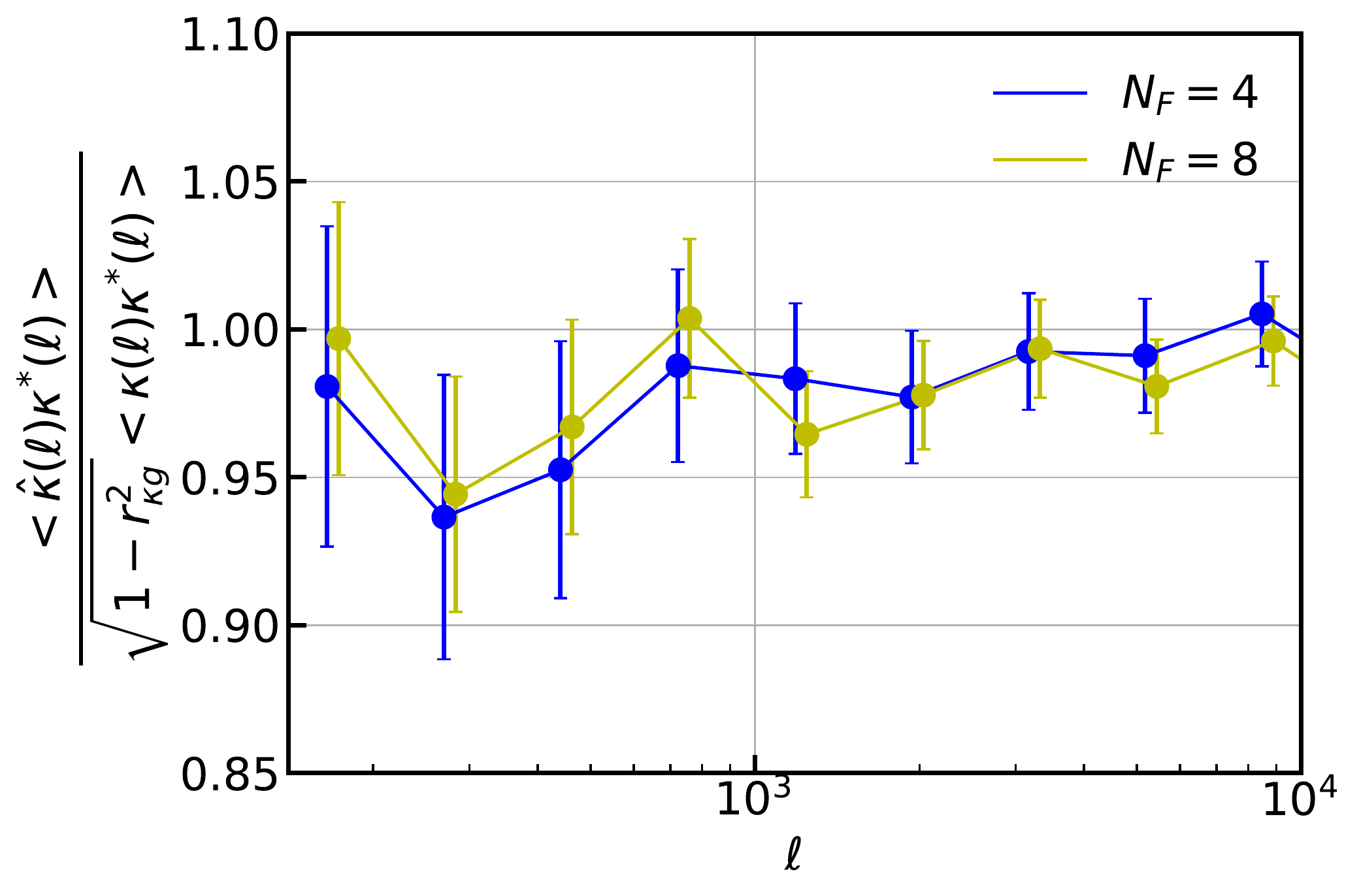}
   \caption{The performance of the Modified ILC method for the combined data when $N_{F}=4$ (the {\it blue line}) and $N_{F}=8$ (the {\it yellow line}). The photo-z range is $0.9<z^{P}<1.1$.}
\label{fig:flux_bin_num}
\end{figure}

We compare the result of four bins in each band to the one with eight bins in Figure \ref{fig:flux_bin_num}.  The galaxies have $0.9<z^{P}<1.1$ and all bands are used.
We only observe a very mild difference between the two cases.
The result from eight bins has slightly smaller statistical uncertainty due to a better sampling of the flux dependence.

\bibliographystyle{raa}
\bibliography{bibtex} 


\label{lastpage}

\end{document}